\documentclass[square]{ws-procs975x65}
\usepackage{graphicx}
\usepackage{color}
\usepackage{epsfig}
\usepackage{dcolumn}
\usepackage{subfigure}
\usepackage{amsmath,amssymb}
\usepackage{array}

\newcommand{\vev}[1]{\left\langle #1 \right\rangle}
\newcommand{\lessim}{\hspace{0.3em}\raisebox{0.4ex}{$<$}\hspace{-0.75em}\raisebox{-.7ex}{$\sim$}\hspace{0.3em}}

\newcommand{\TeV}{\text{TeV}}
\newcommand{\GeV}{\text{GeV}}

\newcommand{\crit}{{\rm crit}}
\newcommand{\tr}{\mbox{tr}}
\newcommand{\gtwo}{I\kern-.1em I\,}
\newcommand{\be}{\begin{equation}}
\newcommand{\ee}{\end{equation}}
\newcommand{\beq}{\begin{eqnarray}}
\newcommand{\eeq}{\end{eqnarray}}
\newcommand{\bpm}{\begin{pmatrix}}
\newcommand{\epm}{\end{pmatrix}}
\newcommand{\cl}{\, \rm C.L.}
\begin{document}

\title{Top Mode pseudo Nambu-Goldstone Boson Higgs Model %
\footnote{This talk is based on \cite{Fukano:2013aea,Fukano:2014zpa,Fukano:2014dta} and given at 
2015 KMI workshop `` Origin of Mass and Strong Coupling Gauge Theories'' (SCGT 15), 
March 3-6, 2015}}
\author{Hidenori S. Fukano\footnote{fukano@kmi.nagoya-u.ac.jp}}
\address{Kobayashi-Maskawa Institute for the Origin of Particles and 
the Universe (KMI) \\ 
 Nagoya University, Nagoya 464-8602, Japan. }

\begin{abstract}
We discuss the Top Mode pseudo Nambu-Goldstone boson Higgs (TMpNGBH) model 
which has recently been proposed as a variant of the top quark condensate model 
in light of the 125 GeV Higgs boson discovered at the LHC. 
In this talk, we focus on the vacuum alignment 
and the phenomenologies of characteristic particles of the TMpNGBH model.
\end{abstract}


%
\section{Introduction}

The ATLAS~\cite{Aad:2012tfa} and CMS collaborations~\cite{Chatrchyan:2012ufa} 
have discovered a $125 \,\GeV$ Standard Model (SM)-like Higgs boson. 
This implies that 
the era to reveal the origin of mass of the elementary particles has come. 
Preceding the discovery of the Higgs boson by about two decades 
the top quark has been discovered at the Tevatron~\cite{Abe:1995hr,Abachi:1995iq}.
The top quark is the heaviest particle among the observed particles 
and its mass is $m_t \simeq 173 \,\GeV$~\cite{Agashe:2014kda}, 
which is coincidentally on the order of the Higgs mass and  
the electroweak symmetry breaking (EWSB) scale 
($v_{_{\rm EW}} \simeq 246\,\GeV$). 

The top quark condensate model~\cite{Miransky:1988xi,Miransky:1989ds,
Bardeen:1989ds} is a scenario 
in which the top quark 
plays a crucial role to explain the dynamical origin for both the EWSB and the Higgs boson. 
However, 
the original top quark condensate model is somewhat far from a realistic situation, 
especially, a Higgs boson predicted as a $t \bar{t}$ bound state 
has the mass in a range of $m_t \lessim m_H \lessim 2 m_t$, 
which cannot be identified with the $125\,\GeV$ Higgs boson at the LHC.

%
\section{Top-Mode pseudo Nambu-Goldstone Boson Higgs (TMpNGBH) model}

Recently,
a variant class of the top quark condensate model,
so-called Top-Mode pseudo Nambu-Goldstone Boson Higgs (TMpNGBH) model, 
was proposed~\cite{Fukano:2013aea,Cheng:2013qwa}. 
In these models 
a composite Higgs boson emerges as 
a pseudo Nambu--Goldstone boson (pNGB) associated with 
the spontaneous breaking of a global symmetry, 
therefore it is light to be identified as the LHC Higgs boson. 

The TMpNGBH model is constructed from 
the top and bottom quarks $q=(t,b)$ and a vectorlike $\chi$ quark, 
a flavor partner of the top quark having 
the same SM charges as those of the right-handed top quark, 
which form a four-fermion interaction:
\beq
{\cal L}_{4f}
=
G_{4f} (\bar{\psi}^i_L \chi_R)(\bar{\chi}_R \psi^i_L)
\,,\label{Lag-4f0}
\eeq
where 
$\psi^i_L \equiv (t_L , b_L, \chi_L)^{T\,i}\,\,( i = 1,2,3)$. 
This four-fermion interaction possesses the global symmetry $G =U(3)_{L} \times U(1)_R$.
When the value of $G_{4f}$ is large enough to form a fermion-bilinear condensate, 
namely $G_{4f}> G_\crit = 8\pi^2/(N_c \Lambda^2)$ 
with $N_c$ being the number of QCD color and $\Lambda$ the cutoff scale of the theory, 
the global symmetry is spontaneously broken down to $H=U(2)_L \times U(1)_V$.
In association with the symmetry breaking, 
the five NGBs emerge as bound states of the $t$ and $\chi$ quarks,  
in addition to a composite heavy scalar boson, corresponding to the $\sigma$ mode of 
the usual Nambu-Jona-Lasinio (NJL) model~\cite{Nambu:1961tp}. 
Three of these five NGBs are eaten by the electroweak gauge bosons 
when the subgroup of $G$ is gauged by the electroweak symmetry 
(and if the condensate is formed in a direction where the electroweak symmetry is broken).   
The other two remain as physical states, 
and they  obtain their masses by additional interaction terms 
which explicitly break 
the global $G=U(3)_{L} \times U(1)_R$ symmetry:
\beq
{\cal L}_h
=
-
\left[\Delta_{\chi \chi} \bar{\chi}_R \chi_L
+ \text{h.c.}
\right]
- G' \left( \bar{\chi}_L \chi_R \right) \left( \bar{\chi}_R \chi_L \right)
\,. \label{Lag-h0}
\eeq 
Then two NGBs become pNGBs, dubbed as top-mode pNGBs (TMpNGBs). 
One of the TMpNGBs, 
which is the $CP$-even scalar ($h^0_t$), 
is identified as the $126\,\GeV$ Higgs boson discovered at the LHC, 
while the other is the $CP$-odd scalar ($A^0_t$).
Furthermore, 
the model includes another four-fermion interaction term,
\beq
{\cal L}_t = 
G'' \left( \bar{\chi}_L \chi_R \right) \left( \bar{t}_R \chi_L\right) + \text{h.c.} 
\,. 
\label{Lag-t}
\eeq
This, combined with 
Eq.(\ref{Lag-4f0}), generates the top quark mass via the top-seesaw mechanism~\cite{
Dobrescu:1997nm,Chivukula:1998wd,He:2001fz,Fukano:2012qx,Fukano:2013kia}. 

Note that Eq.(\ref{Lag-t}) also explicitly breaks the $G$-symmetry, 
but does not contribute to the TMpNGBs' masses ($m_{h^0_t}$ and $m_{A^0_t}$) 
at the leading order.     
However, it was shown that at the next-to-leading order, 
the term in Eq.(\ref{Lag-t}) gives large corrections to the masses 
of $h^0_t$ and $A^0_t$ 
via the top and $\chi$-quark loops~\cite{Fukano:2013aea}.
This, 
namely the fact that even a small explicit breaking term causes 
large correction to physical quantities at the loop level, 
poses a question: 
is the vacuum alignment stable at the loop level ? 
We address this question based on   
an effective Lagrangian described by the TMpNGBs ($h^0_t$ and $A^0_t$), 
the $t'$ quark, the SM gauge bosons and fermions, including terms explicitly breaking 
the global $U(3)_L \times U(1)_R$ symmetry.
%

\section{Vacuum Alignment of TMpNGBH model}
\label{sec-vacuum-alignment} 

The effective Lagrangian relevant for the vacuum alignment is given by
\beq
{\cal L}_{\rm eff}(U)
&=&
\frac{f^2}{2}
\tr \left[ D_\mu U^\dagger D^\mu U \Sigma_0 \right]
-
\tilde{m}_{\chi}
\left[
\bar{\psi}_L {\cal M}_f(U) \psi_R + \text{h.c.}
\right]
\nonumber\\
&&
-
c_1f^2
\tr\left[
U^\dagger \Sigma_0
U \Sigma_0
\right]
+
c_2 f^2
\tr\left[
U \Sigma_0
+
\Sigma_0 U^\dagger
\right]
\,,\label{start-NLsM}
\eeq 
where the unitary matrix $U$ parameterizes the five NGBs and is given by
\beq
U &=& 
\exp \left[
\frac{i}{f} \left( 
\sum_{a=4,5,6,7} 
\pi^a_t \lambda^a 
+ \pi^A_t \Sigma_0 \right) 
\right] 
\,.
\label{eq:defU}
\eeq
Here, $f$ is a decay constant, 
the Gell-Mann matrices $\lambda^a$ are normalized as 
${\rm tr}[\lambda^a \lambda^b]=2 \delta^{ab}$, and $\Sigma_0$ is defined 
as $\Sigma_0 \equiv \text{diag}(0,0,1)$. 
$D_\mu U =\left( \partial_\mu - i g \hat{W}_\mu + i g' \hat{B}_\mu \right) U$,
$\hat{W}_\mu = \sum^3_{\hat{a}=1} W^{\hat{a}}_\mu (\lambda^{\hat{a}}/2)$,
$\hat{B}_\mu = B_\mu \cdot\text{diag}(1/2,1/2,0)$, 
$W_\mu$ and $B_\mu$ are the usual $SU(2)_L$ and $U(1)_Y$ gauge fields 
with gauge couplings $g$ and $g'$, respectively.
${\cal M}_f(U),\tilde{m}_\chi$ are given by
\beq
{\cal M}_f(U)
=
U \Sigma_0
+
\frac{G''}{G_{4f}} \Sigma_0 U 
\bpm
0 & 0 & 0 \\
0 & 0 & 0 \\
1 & 0 & 0 \\
\epm
\,,
\tilde{m}_{\chi} 
= 
\frac{1}{\sqrt{2}} y f 
= \sqrt{\frac{8 \pi^2}{N_c \ln (\Lambda^2/\Lambda^2_\chi)}} f 
\,,
\label{m-chi}
\eeq
where $\Lambda$ is the cutoff scale of the ultraviolet theory and 
$\Lambda_\chi$ is an infrared scale corresponding to 
the cutoff scale of the effective theory Eq.(\ref{start-NLsM}).
The coefficients $c_{1}$ and $c_{2}$ in Eq.(\ref{start-NLsM}) are given by 
\beq
c_1 
= 
\frac{y^2}{2} \frac{G'}{G^2_{4f}} 
\,,\quad
c_2
=
\frac{y }{\sqrt{2}f} \frac{\Delta_{\chi \chi}}{G_{4f}}
\,.
\label{def-c1-c2}
\eeq 

At the tree level, the form of the potential term for NGBs, 
corresponding to the second line of Eq.(\ref{start-NLsM}), 
is determined solely by the ${\cal L}_h$. 
The effect of the explicit breaking terms in ${\cal L}_t$ and the electroweak sector appear 
only at loop level. 
Therefore, to see the effect of all the explicit breaking terms, 
we compute the effective Lagrangian at one-loop level 
by including all the contributions from the NGBs, electroweak gauge bosons, as well as fermions. 
The effective Lagrangian is calculated by keeping only the quadratic divergent terms, 
and the resultant expression becomes as follows 
(for the detail of the calculation, see \cite{Fukano:2014dta}):
\beq
{\cal L}^{\text{1-loop}}_{\rm eff} (U)
= 
\frac{F^2}{2}
\tr
\left[
D_\mu U^\dagger D^\mu U 
\Sigma_0
\right]
-
\tilde{m}_\chi
\left[
\bar{\psi}_L {\cal M}_f(U) \psi_R + \text{h.c.}
\right]
-
V_{\rm eff} (U) 
\,,\label{1loop-eff-Lag-reno}
\eeq 
where 
the effective potential $V_{\rm eff} (U)$ is given by
\beq
V_{\rm eff} (U) 
=
C_1 F^2 
\tr\left[
U^\dagger \Sigma_0
U \Sigma_0
\right]
-
C_2 F^2 
\tr\left[
U \Sigma_0
+
\Sigma_0 U^\dagger
\right]
\,.\label{eff-potential-TMP}
\eeq
The quadratic divergences 
can be absorbed by redefinitions of the bare coupling $f$, $c_1$ and $c_2$: 
\beq
F^2
&=&
f^2 - \frac{\Lambda^2_\chi}{4\pi^2} 
=  
\frac{N_c}{8\pi^2} \tilde{m}^2_\chi \ln \frac{\Lambda^2}{\Lambda^2_\chi} 
- \frac{\Lambda^2_\chi}{4 \pi^2} 
\,,\label{redef-decayconstant} 
\\ 
C_1 F^2 &=&
c_1 f^2
\left( 1 - \frac{3\Lambda^2_\chi}{8\pi^2 f^2}\right) 
-
\frac{f^2 \Lambda^2_\chi}{32\pi^2} 
\left(
 \frac{9}{4} g^2 + \frac{3}{4} g'^2 
+
2N_c y^2 \left( \frac{G''}{G_{4f}}\right)^2
\right) 
\,, \label{redef-c1} \\ 
C_2F^2 
&=& 
c_2 f^2
\left( 1 - \frac{5 \Lambda^2_\chi }{32\pi^2 f^2}\right) 
\,. \label{redef-c2}
\eeq 

Let us address the vacuum alignment of the TMpNGBH model 
based on the effective potential Eq.(\ref{eff-potential-TMP}). 
First, 
with appropriate chiral $U(3)_{L,R}$ rotations of fermion fields $\psi_{L,R}$ 
and redefinition of the $\Delta_{\chi\chi}$, 
we parameterize the vacuum expectation value of $U$ 
by a single angle parameter $\theta$ as 
\beq
\vev{U} 
=
\bpm
\cos \theta & 0 & \sin \theta \\[1ex]
0 & 1 & 0 \\[1ex]
-\sin \theta & 0 &  \cos \theta
\epm
\,. \label{def-vev-U}
\eeq
Taking $U = \vev{U}$, 
we have 
$V_{\rm eff}(\vev{U}) = 
F^2 \left[ C_1 \cdot \cos^2\theta - 2 C_2 \cdot \cos\theta \right] $.
It is possible to determine the vacuum alignment by minimizing the above potential energy 
with respect to the alignment parameter $\cos\theta$. 
In the present model, 
we find that the potential energy $V_{\rm eff}(\vev{U})$
is minimized at a nonzero $\theta=\theta_h$ with 
\beq
\cos\theta \Bigg|_{\theta = \theta_h} = \frac{C_2}{C_1} 
\quad \quad
\text{ only if \quad $C_1 >0$ and $\left| \dfrac{C_2}{C_1} \right | < 1$}
\,,\label{V-min-TMP}
\eeq
to realize the desired vacuum in which 
the electroweak symmetry is broken.  

From the effective potential, 
we find the non-vanishing elements of the NGB mass-squared matrix take the following forms:
\beq
\bpm
m^2_{44} & m^2_{4A} \\[1ex] m^2_{A4} & m^2_{AA}
\epm
=
2 C_1
\times 
\bpm
\cos\theta_h & -\sin\theta_h \\[1ex]
\sin\theta_h & \cos\theta_h
\epm
\bpm
0 & 0 \\[1ex] 
0 & 1
\epm
\bpm
\cos\theta_h & \sin\theta_h \\[1ex]
-\sin\theta_h & \cos\theta_h
\epm
\,,\label{44-AA-mass-2}
\eeq
and
\beq
m^2_{55} 
= 
2 C_1  \sin^2\theta_h
\,.\label{55-mass-2}
\eeq 
Note that the stability of the effective potential requires $C_1 \geq 0$~\cite{Fukano:2014dta}. 
The massive state in Eq.(\ref{44-AA-mass-2}) is identified as the $CP$-odd scalar $A^0_t$ 
($A^0_t\equiv -\pi^4_t \sin\theta_h + \pi^A_t \cos\theta_h$), 
while that in Eq.(\ref{55-mass-2}) is the $CP$-even scalar ($\pi_5\equiv h^0_t$), 
dubbed as the ``tHiggs".
These masses are related by the alignment parameter $\theta_h$:     
\beq
m^2_{A^0_t} 
&=&
2C_1  
\,,\label{CPodd-TMP-mass}
\\[1ex]
m^2_{h^0_t} 
&=&
2C_1   \sin^2\theta_h 
\nonumber\\
&=& 
m^2_{A^0_t} \sin^2\theta_h
\,.\label{CPeven-TMP-mass}
\eeq 
Other three eigenvalues of mass-squared matrix vanish, which corresponds to 
three massless NGBs ($\pi^{6,7}_t\,,\pi^4_t \cos\theta_h + \pi^A_t \sin\theta_h$).  
These are the would-be NGBs to be eaten by the electroweak gauge bosons.  
It should be noted from Eqs.(\ref{CPodd-TMP-mass}) and (\ref{CPeven-TMP-mass}) 
that the quadratic divergent contributions to masses of TMpNGBs have been 
fully absorbed into the renormalization of the decay constant $F$, 
the coefficient $C_1$ and the alignment parameter $\theta_h$ (or the coefficient $C_2$).

\section{Implications for collider physics}
\label{sec-phenomenologies-A-tprime}

In this section, 
we discuss phenomenological implications for the TMpNGBH model. 
We take the alignment parameter $\cos\theta_h$ 
in the range of $0.97 \leq \cos \theta_h \lessim1$.
This is the range where 
the coupling property of the tHiggs to SM particles is consistent with the LHC data 
at $95\%\cl$~\cite{Fukano:2014zpa}.
For $0.97 \le \cos\theta_h \lesssim 1$ 
the masses of $A^0_t$ and $t'$ monotonically increase from 
$(m_{A^0_t}, m_{t'})=(518\,\GeV, 1.85\,\TeV)$ 
to infinity as $\cos\theta_h \to 1$. 
This value of $m_{t'}$ is consistent with the electroweak precision tests~\cite{ %
Peskin:1990zt,Peskin:1991sw} as shown in \cite{Fukano:2013aea}.
We thus study the LHC phenomenologies of $A^0_t$ and $t'$ 
with their masses from $(m_{A^0_t}, m_{t'}) =(518 \,\GeV, 1.85 \,\TeV)$ 
to certain heavier mass regions which are considered to be relevant to the LHC.

\begin{figure}[h]
\begin{center}
\begin{tabular}{cc}
{
\begin{minipage}[t]{0.4\textwidth}
\includegraphics[scale=0.62]{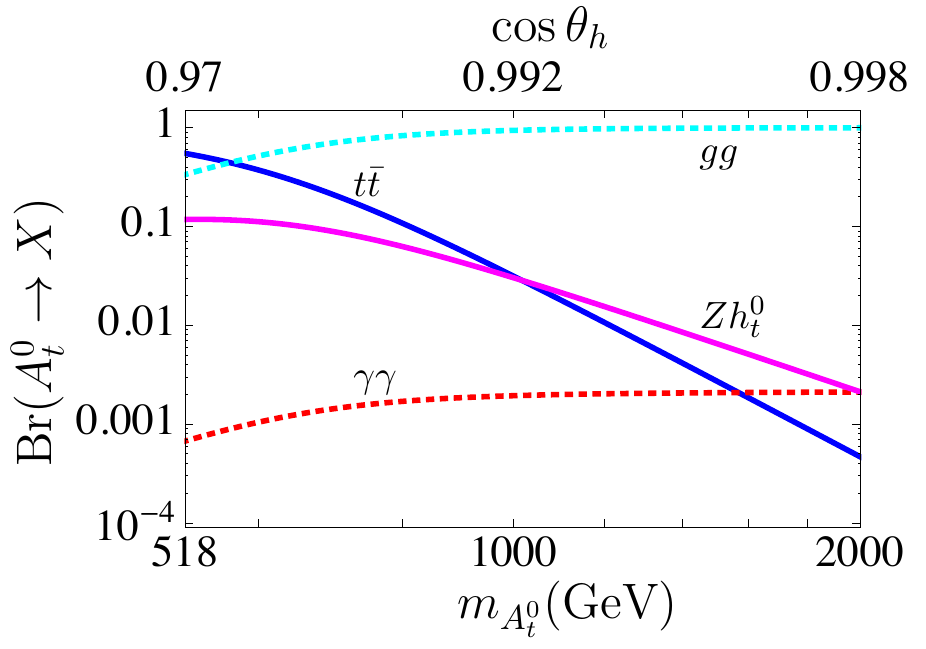} 
\end{minipage}
}
&
{
\hspace{15mm}
\begin{minipage}[t]{0.4\textwidth}
\includegraphics[scale=0.58]{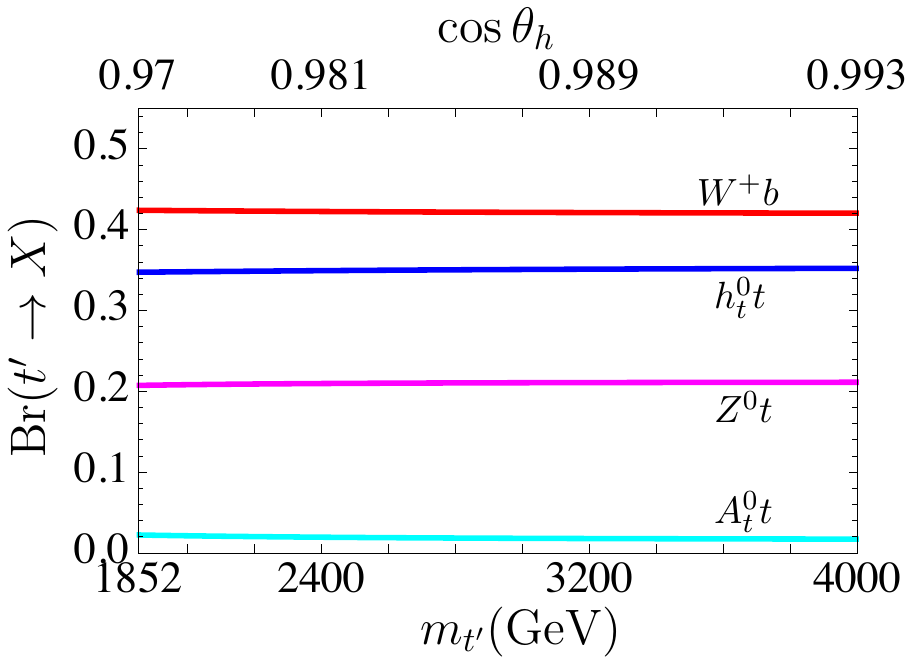} 
\end{minipage}
}
\end{tabular}
\caption[]{
The branching ratios of $A^0_t$ (left panel) and $t'$ (right panel) 
as functions of $m_{A^0_t}$ and $m_{t'}$, respectively. 
Values of $\cos \theta_h$ 
are also shown in the upper horizontal axes. 
\label{Br-A-tprime}}
\end{center}
\end{figure}

The couplings of $A^0_t$ to the SM particles, 
the tHiggs ($h^0_t$) and the $t'$ quark can be read off 
from the Lagrangian Eq.(\ref{1loop-eff-Lag-reno}). 
The explicit expressions of the  partial decay widths relevant to the LHC study 
can be found in \cite{Fukano:2014zpa} with the replacement,  
$f \to F$ and $\theta \to \theta_h$. 
In Fig.~\ref{Br-A-tprime}, 
we plot the branching ratio of $A^0_t$ as a function of $m_{A^0_t}$ 
in the range of $518 \,\GeV \leq m_{A^0_t} \leq 2\,\TeV$ 
in the left panel of Fig.~\ref{Br-A-tprime}. 
In this plot, we also indicate the corresponding values of $\cos\theta_h$ 
in the upper horizontal axis. 
From the plot we see that, in the smaller mass region, 
the $t\bar{t}$ and $gg$ modes 
are the dominant decay channels, 
and therefore the main production process is the gluon-gluon fusion (ggF). 
The 8 TeV LHC cross sections 
$pp \to A^0_t \to gg/tt$ for $m_{A^0_t} \geq 1\,\TeV$ 
have not seriously been limited by the currently available data yet.
It is therefore to be expected that more data from the upcoming Run-II would probe the $A^0_t$ 
through these channels. 
Another interesting channel would be $A^0_t \to Z h^0_t$. 
However, with the updated branching ratio, 
this channel seems to be rather challenging even 
at the $\sqrt{s}=14\,\TeV$ LHC with 3000 ${\rm fb}^{-1}$ data 
due to the small branching ratio in the smaller mass region.

The $t'$ quark arises as a mixture of the gauge-eigenstate top and $\chi$-quarks 
through the diagonalization of the fermion mass matrix 
in the effective Lagrangian Eq.(\ref{1loop-eff-Lag-reno}).   
The explicit expressions of the $t'$ couplings and the partial decay widths 
relevant to the LHC study are listed in \cite{Fukano:2014dta}. 
In the right panel of Fig.~\ref{Br-A-tprime}, 
we plot the branching ratios of the $t'$ quark as a function of $m_{t'}$. 
In the same way as the plot for the branching ratios of $A^0_t$, 
the corresponding value of $\cos\theta_h$ is also shown in the upper horizontal axis.
From the figure we read off $\text{Br}(t' \to W^+ b) \simeq \text{Br}(t' \to h^0_t t) \simeq 0.4$,  
$\text{Br}(t' \to Zt) \simeq 0.21$ and 
$   \text{Br}(t' \to A^0_t t) \simeq 0.02$.
It is worth comparing these values with the branching ratios of the ``singlet $t'$ quark" 
in a benchmark model of $t'$ quark~\cite{delAguila:1989rq}, 
$\text{Br}(t' \to W^+ b) \simeq 0.5,  
\text{Br}(t' \to Zt) \simeq 0.25, 
\text{Br}(t' \to h t) \simeq 0.25$, 
for $m_{t'} \simeq 2\,\TeV$~\cite{AguilarSaavedra:2009es,Aguilar-Saavedra:2013qpa}.
It is interesting to note that 
$\text{Br}(t' \to h^0_t t)$ in the present model 
is by about $40\,\%$ larger than that in the benchmark model. 
This is essentially 
due to the large $ht't$ coupling, 
which is the very consequence of the top quark condensate scenario.   

\section{Summary}
\label{summary}

We presented the Top Mode pseudo Nambu-Goldstone boson Higgs (TMpNGBH) model 
which has recently been proposed as a variant of the top quark condensate model 
in light of the 125 GeV Higgs boson discovered at the LHC. 
We also discussed the vacuum alignment problem of TMpNGBH model
based on the one-loop effective Lagrangian for the NGB sector, 
taking into account all the explicit breaking effects, 
including electroweak gauge interactions and four fermion interactions 
responsible for the top-seesaw mechanism. 
We found that the correct vacuum is determined by 
the configuration which minimizes the one-loop effective potential. 
It was found that the true vacuum is parameterized by $\cos\theta_h$, 
and a non-zero value of $\cos\theta_h$ realizes the EWSB phase 
with the appropriate breaking scale. 
Furthermore,
we also discussed the phenomenological implications of the TMpNGBH model  
on the vacuum aligned at the one-loop level.

%

\bibliography{ref-alignment,ref-BFM,ref-NLsM,ref-chiralsymmetry,ref-compositeHiggs,ref-SMHiggs,ref-EWPT,ref-2HDM,ref-TC,ref-TMSM,ref-PDG-PDF,ref-Higgs-ATLAS,ref-Higgs-CMS,ref-LHC-woSMHiggs,ref-VLQ,ref-top-quark}
\end{document}